\documentclass[sigconf,natbib=true]{acmart}

\usepackage{natbib}
\usepackage{multirow}
\usepackage{balance} 
\title{Query-dominant User Interest Network for Large-Scale Search Ranking}

\author{Tong Guo}
\affiliation{%
  \institution{Kuaishou Technology Co., Ltd.}
  \city{Beijing}
  \country{China}
}
\email{guotong03@kuaishou.com}
\author{Xuanping Li}
\authornote{corresponding author}
\affiliation{%
  \institution{Kuaishou Technology Co., Ltd.}
  \city{Beijing}
  \country{China}
}

\email{lixuanping@kuaishou.com}
\author{Haitao Yang}
\author{Xiao Liang}
\affiliation{%
  \institution{Kuaishou Technology Co., Ltd.}
  \city{Beijing}
  \country{China}
}
\email{yanghaitao@kuaishou.com}
\email{liangxiao@kuaishou.com}
\author{Yong Yuan}
\author{Jingyou Hou}
\affiliation{%
  \institution{Kuaishou Technology Co., Ltd.}
  \city{Beijing}
  \country{China}
}
\email{yuanyong@kuaishou.com}
\email{houjingyou@kuaishou.com}
\author{Bingqing Ke}
\author{Chao Zhang}
\affiliation{%
  \institution{Kuaishou Technology Co., Ltd.}
  \city{Beijing}
  \country{China}
}
\email{kebingqing@kuaishou.com}
\email{zhangchao@kuaishou.com}
\author{Junlin He}
\author{Shunyu Zhang}
\affiliation{%
  \institution{Kuaishou Technology Co., Ltd.}
  \city{Beijing}
  \country{China}
}
\email{hejunlin@kuaishou.com}
\email{zhangshunyu@kuaishou.com}

\author{Enyun Yu}
\affiliation{%
  \institution{unaffiliated}
  \country{China}
}
\email{yuenyun@126.com}
\author{Wenwu Ou}
\affiliation{%
  \institution{unaffiliated}
  \country{China}
}
\email{ouwenwu@gmail.com}

\copyrightyear{2023}
\acmYear{2023}
\setcopyright{acmlicensed}
\acmConference[CIKM '23] {Proceedings of the 32nd ACM International Conference on Information and Knowledge Management}{October 21--25, 2023}{Birmingham, United Kingdom.}
\acmBooktitle{Proceedings of the 32nd ACM International Conference on Information and Knowledge Management (CIKM '23), October 21--25, 2023, Birmingham, United Kingdom}
\acmPrice{15.00}
\acmISBN{979-8-4007-0124-5/23/10}
\acmDOI{10.1145/3583780.3615022}

\begin{document}
\begin{abstract}

Historical behaviors have shown great effect and potential in various prediction tasks, including recommendation and information retrieval. The overall historical behaviors are various but noisy while search behaviors are always sparse. Most existing approaches in personalized search ranking adopt the sparse search behaviors to learn representation with bottleneck, which do not sufficiently exploit the crucial long-term interest. In fact, there is no doubt that user long-term interest is various but noisy for instant search, and how to exploit it well still remains an open problem.

To tackle this problem, in this work, we propose a novel model named \textbf{Q}uery-dominant user \textbf{I}nterest \textbf{N}etwork (\textbf{QIN}), including two cascade units to filter the raw user behaviors and reweigh the behavior subsequences. Specifically, we propose a relevance search unit (\textbf{RSU}), which aims to search a subsequence relevant to the query first and then search the sub-subsequences relevant to the target item. These items are then fed into an attention unit called Fused Attention Unit (\textbf{FAU}). It should be able to calculate attention scores from the ID field and attribute field separately, and then adaptively fuse the item embedding and content embedding based on the user engagement of past period.
Extensive experiments and ablation studies on real-world datasets demonstrate the superiority of our model over state-of-the-art methods. The QIN now has been successfully deployed on Kuaishou search, an online video search platform, and obtained 7.6\% improvement on CTR. 

\end{abstract}




\maketitle

\section{Introduction}

Personalized Search Ranking (PSR) system plays an important role in online e-commerce \cite{ai2017learning} and content-sharing platforms \cite{eksombatchai2018pixie}. Users describe their needs through queries to get desired results with the help of the PSR system. Users interact with the items and have a joyful time through the platforms, calling user engagement. 
To improve user engagement, search engines should sincerely consider not only consider the query-item relevance but also take into account users' preferences.

\begin{figure}[t]
  \centering
  \includegraphics[width=\linewidth]{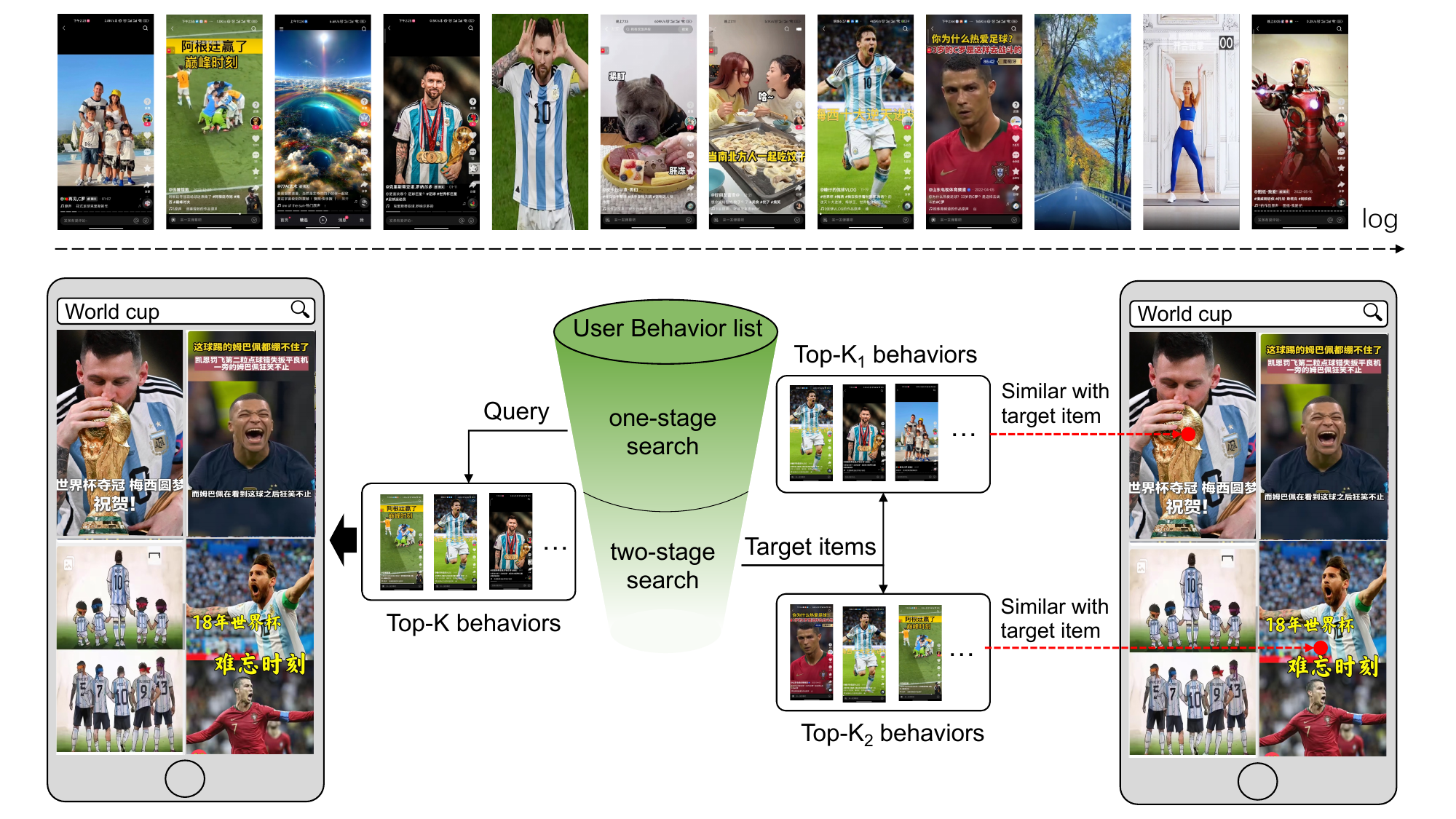}
  \caption{The illustration of the Relevance Search Unit. The first stage of RSU exploits query embedding to retrieve Top-$K_1$ relevant items from the redundant historical behaviors. On this basis, the second stage needs to find the most relevant Top-$K_2$ items in the subsequences retrieved from the result of the first stage for each target items. The two-stage RSU can effectively filter the irrelevant items relevant to the query and target item.}
  \label{fig:example}
  \Description{An inverted cone called RSU is located in the center of the picture. User-watching log is fed into it through one-stage search and two-stage search. The first-stage search unit uses query embedding to obtain top-k items as user interest sequences to model user preference, and the second-stage search unit additionally exploits target item embedding to obtain the subsequences of the top-k relevant items. One-stage user subsequences vary by query, while two-stage user subsequences vary by item.}
\end{figure}
To address this issue, many previous methods\cite{ai2019zero,ai2017learning,bennett2012modeling, bi2020transformer,dumais2016personalized} have introduced historical search behaviors into their user interest networks \cite{covington2016deep}, and extracted behavioral patterns for recommendation purposes\cite{cheng2022ihgnn}.
Recently, modeling long-term user interests has become an increasingly popular trend in the recommender system community \cite{pi2020search, tian2022multi, zhou2019deep}. However, few works have explored such line of study in search engines. We argue that by effectively and efficiently taking into account the overall users' historical behaviors not just the search history, search results can be greatly improved by personalization. For example, a user  watched many videos of Messi on the platform while never searched for the query "Messi". When he initiates a search request, the top videos retrieved by keywords about \emph{football} should be related to Messi other than other athletes.

To this end, noticing the success of user long-term interest exploration for sequential recommendation \cite{pi2020search}, we wish to take advantage of it in search scenarios. The general framework of sequential recommendation can be summarized as the process of extracting and purifying useful behaviors and attributes. This is non-trivial to transfer it to the PSR task, facing two challenges as follows:
\begin{itemize}
\item {\textbf{First}}: We view all the interactions between users and recommendation/search results as the overall behaviors. The users' overall behaviors are always irrelevant to the search query while the users' search behaviors are relevant but sparse. Directly following the sequential recommendation task \cite{pi2020search, zhou2018deep}, the overall behaviors will model the users' interest which is irrelevant to the query. The search behaviors are so sparse that the interaction lists lack much information about the users' long-term interests. Therefore, rough utilization might lead to negative effects and how to exploit users' overall behaviors in this scene remains challenging.
\item {\textbf{Second}}: The behaviors in recommendation can be formed as (user, item) pair while (user, query, item) in search. Users' historical behaviors would be very sparse for the given query, resulting in a degraded performance with only the ID field as input. Paying attention to the attribute field to enhance the ID field is crucial. Besides, similar items, with different depths of user engagement, should have different contributions to prediction, but the existing methods treat them equally.
\end{itemize}

To address these challenges, the QIN method is proposed with two cascade units --- relevance search unit (\textbf{RSU}) to filter the irrelevant behaviors and fused attention unit (\textbf{FAU}) to model query-dominant user interests. Specially, the RSU aims to obtain subsequences of users' overall behaviors to capture the long-term interest. We show the roadmap in Figure \ref{fig:example}. The users' overall behaviors are fed into search units to seek top-K relevant items. The RSU includes two stages: the first stage searches a subsequence relevant to the query from user's behavior list and the second stage searches sub-subsequences relevant to the target items from the result of the first stage .
Then FAU decouples the calculation of attention score with the ID list and attribute list, and the attention scores of two field are fused dynamically according to the
depth of user engagement. The deeper the user's engagement with the item, the higher the attention score should be. For example, among the user-watching videos in history, a video that has been played for 100 seconds should have a higher weight than a video that has only been played for 1 second.
By combining these two modules, QIN extracts and refines useful attributes from overall user behaviors, leading to more personalized and satisfied search results and obtained 7.6\% improvement on CTR.

In summary, the main contributions of this paper are briefly outlined as follows:
\begin{itemize}
    \item We highlight the significance of utilizing user behaviors in personalized search ranking, and demonstrate it can greatly boost user engagement if search results are relevant and based on user's historical preferences. To overcome the practical challenges of search behavior sparsity and long-tail queries, we take into account user long-term behaviors of both search and recommendation, and carefully design the proposed QIN method.
    \item Specifically, QIN utilizes user long-term behaviors with a novel two-stage attention module: Search Relevant Unit(RSU) and Fused Attention Unit(FAU). RSU extracts the relevant subsequence from the users' overall behaviors to alleviate the noise in overall behaviors. FAU complements the ID feature with the content feature to model the user interest from sequence data. Then the attention scores of both fields are fused with user engagement.
    \item We conduct extensive experiments on real-world datasets and empirical results demonstrate the effectiveness of our method QIN. QIN has also outperformed our highly-optimized production baseline, and has been launched to serve hundreds of millions users on Kuaishou search. 

\end{itemize}

\section{Related Work}
\subsection{Personalized Search Ranking}
Recently, personalized search has been an appealing topic in the community, and is widely applied in real-world application \cite{agichtein2006learning, grbovic2018real, ye2021don}. We roughly split them into two lines. On the one hand, graph-based methods focused on expanding the potential interest of users\cite{ai2019explainable, Han2023}. IHGNN \cite{cheng2022ihgnn} proposed to enhance the item embedding from the collected signal from the triple interactive hypergraph. On the other hand, attention-based methods capture the pattern of user interest according to the historical behaviors\cite{bi2021learning,kocayusufoglu2022multi}. Guo at al. \cite{guo2019attentive} exploited the attention mechanism on both users' short and long-term preferences for personalized product search. Both kinds of methods extract the user preferences from either the potential expansion set of user interest or the natural behavior set. However, there exists a limitation in that they only utilize the user's historical search behavior to model the user's preferences. Our proposed model significantly expands the set of user behaviors by searching the complete user history behavior.
\subsection{User Interest Model}
A series of works focus on learning the representation of user interest from historical behaviors, adopting different architecture like DNNs \cite{zhou2018deep}, CNNs \cite{tang2018personalized}, RNNs \cite{hidasi2015session} and transformer \cite{sun2019bert4rec}. Due to the great power of attention, researchers adopt this architecture for modeling user interest. SASRec \cite{kang2018self} demonstrated the effectiveness of self-attention with the SOTA performance. DIF-SR \cite{xie2022decoupled} raised that early integration of side information can lead to a rank bottleneck of the attention matrices and decouples attention calculation in \cite{liu2021noninvasive} to avoid being bounded by head projection size of the attention matrices. We propose a novel attention mechanism to model the query-dominant user interest, and prove the superiority with the ablation study. Authors in \cite{zheng2022disentangling} tried to disentangle long and short-term interest for recommendation and model the long and short-term interest separately. We introduce the approach to model the long-term interest for search ranking and prove the effectiveness of the proposed method.

\section{PRELIMINARIES}
In this section, we first formulate the personalized search ranking task, and discuss the bottleneck of the recent PSR methods.
\subsection{Task Formulation}
Suppose we have a search engine with a user set $\{u\mid u\in\mathcal{U}\}$, an item set $\{i\mid i\in\mathcal{I}\}$, a possible query set $\{q\mid q\in\mathcal{Q}\}$, historical behaviors $\mathcal{B}$. Each item $\{i\mid i\in\mathcal{B}\}$ indicates that user $u$ ever interacted with it. Depth of user engagement can be either a numerical feature like rating or a dense feature like watch time. $R$ means the look-up embedding, such as $R^{(ED)}$ that means the embedding of user engagement depth. We use $y_{uqi}$ indicates whether the ($u$,$q$,$i$) interaction happens. The goal of our model is to predict the probability of a user $u$ interacting with an item $i$ when searching query $q$. Given the whole items set $\mathcal{I}$, the overall historical behaviors $\mathcal{B}$ and search behaviors $\mathcal{B}_q$, their relationship is shown in Figure \ref{fig:bottleneck} (a). 
\subsection{Inherent Bottleneck in Search Behavior \label{sec:bottle}}

\begin{figure}[t]
  \centering
  \includegraphics[width=\linewidth]{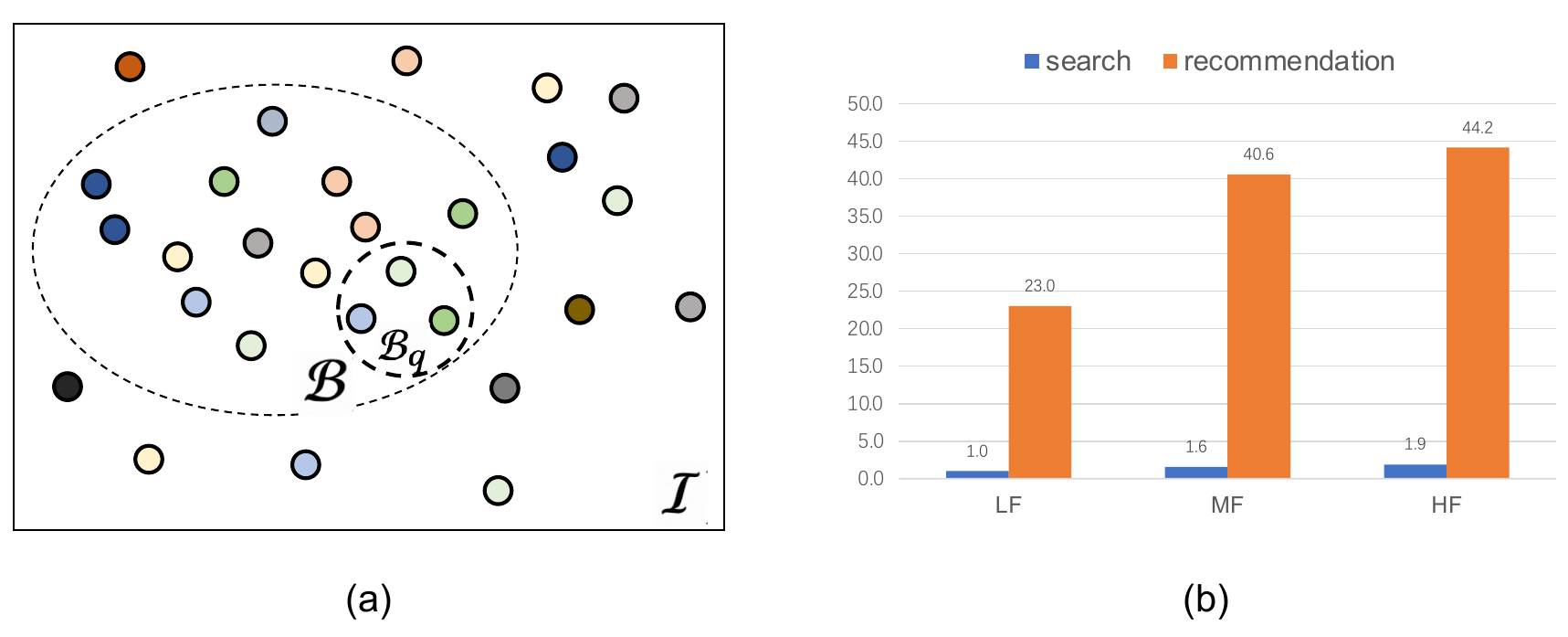}
  \caption{(a) gives relationship among $\mathcal{I,B},\mathcal{B}_q$, (b) shows the average behavior sequence length of search and recommendation in real world in low frequency (LF), medium frequency (MF) and high frequency (HF). }
  \label{fig:bottleneck}
  \Description{}
\end{figure}
\begin{figure*}[ht]
   \centering
  \includegraphics[width=\linewidth]{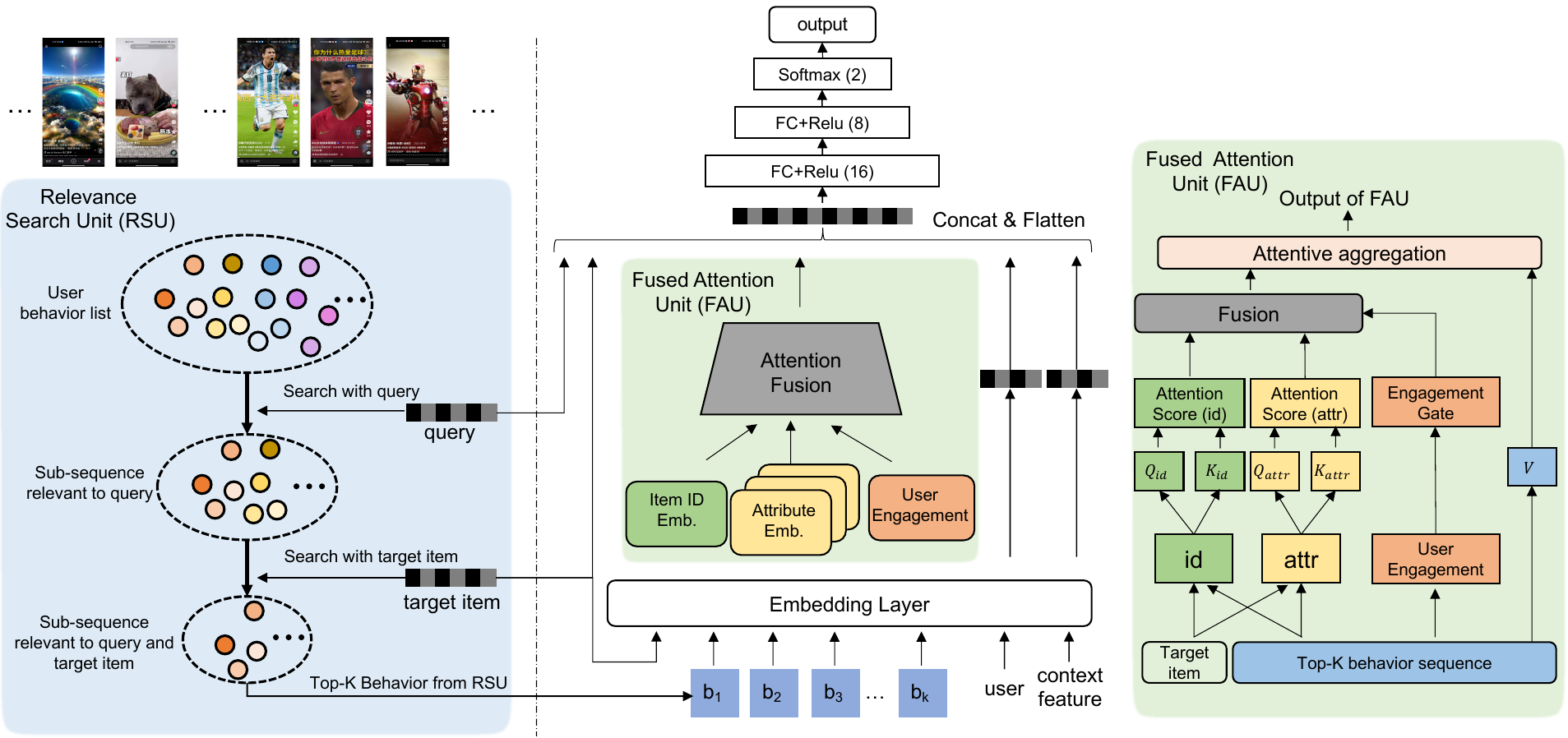}
  \caption{Overview of the proposed model, including the relevance search unit and fused attention unit. The raw historical behavior list is fed into the RSU to select K behaviors. The FAU aims to model the user interest from the subsequences, and fuse the attention scores of ID field and attribute field with user engagement. We stack two FAUs where one is self-attention for enhancement of the representation of historical items and the other one is target-attention for enhancement of the representation of the target item.}
  \label{fig:framework}
  \Description{}
\end{figure*}

Search is a process of obtaining information for users. In this process, most of the queries are never searched by users in their behavior history. The historical behaviors are usually so sparse and limited to capture the user long-term interest \cite{matthijs2011personalizing}. 

Real-world data analysis is performed to explore the historical behaviors of users in search. We grabbed all the search behaviors from Kuaishou in one week and track their overall behaviors through the platform. We divide the users into high frequency, medium frequency, and low frequency according to user activity level and calculate the average sequence length, e.g., the length of video watch list in history. The values of the vertical axis are desensitized by setting the average length of historical behaviors in LF user group to 1. As Figure \ref{fig:bottleneck} (b) shows, the average gap between search and recommendation reaches 20 times, which means users' search behaviors are very sparse compared with recommendation.

The former analysis demonstrates the bottleneck of search behaviors, so that it's hard to truly reflect the user's long-term interest by using the representation only learned from the search behaviors. An intuitive approach is to utilize the overall historical behaviors to enhance the interest representation of search users. But the long behavior list may impose a negative effect on the search ranking task because of the irrelevance. An effective solution to this shortcoming should be raised for the extraction of long-term interest.

\section{METHODOLOGY}
\subsection{Relevance Search Unit}
As we have analyzed in Section \ref{sec:bottle}, historical behaviors are too noisy for the query. To alleviate this, we utilize Relevance Search Unit (RSU) to seek top-K relevant items from long-term historical behaviors. Give a query $q$ and user behaviors list $\mathcal{B} = [b_1,b_2,\cdots,b_T]$, the RSU calculates relevance score $r_{b_t}^q$ for each behavior $b_t$ with the query $q$, and then select the top-K behaviors. These subsequences expand the set of search behavior without imposing negative impact upon relevance. For the given query $q$, if all the target items share the same subsequences, it will result in low distinguishability in one search session. The user interest, that determine whether a user will interact with an item, should be captured from historical behavior relevant to the target item. Thus we further exploit another search unit from the target item perspective.
Given the target item $i$, the top-$K_2$ sub-subsequences can be retrieved from the top-$K_1$ result $\mathcal{B}^q$. The first stage search unit expands the potential set of user interest for a global user pattern regarding query and the second stage search unit focuses on the local pattern regarding various target items.

\textbf{Relevance Calculation.} To bridge the gap between the representation space of the queries and items, as different types of entities in different domains, we need to project them into the same space before calculating the similarity. 
Hence we adopt the pre-trained embedding latent space to catch the relevance between entities. From a personalized perspective, users express different potential demands even from the same query. Therefore, it is possible to match the semantics of the query and the text field of the item with the support of advanced methods of pre-trained models that have seen large-scale data. We generate the pre-trained embeddings with the proposed model in \cite{devlin2018bert}. Furthermore, an upgraded version of semantics is proposed for multi-modal retrieval on the multimedia content platform. It's from a advanced model in \cite{wang2022modality} for online A/B testing. These two both use the advanced representation methods to auxiliary the search ranking.

\textbf{Search Process.} With the relevance score, we then introduce the search stages. Considering the noise and the excessive diversity in the behavior list, the first stage calculates the relevance scores between the query and behaviors, which is supposed to filter the majority of irrelevant historical behaviors. The results of search unit describe the user's long-term interest corresponding to the query. The behavior subsequence contains behaviors with different relevant granularities, so it greatly increases the sequence length comparing with the search sequence. It's also intuitive that the user's historical behaviors relevant to the \emph{dress} can enrich the behavior history of the \emph{yellow floral dress}. 

Moreover, the lack of discrepancy across the candidate items inspires us to further utilize the target item for another search process. This search unit leverages the result of the prior search unit to sample top-K relevant items relevant to the target item. Specifically, we first use the query to retrieve the top-$K_1$ items from the complete behavior sequence and then use the target item to retrieve the top-$K_2$ items from the $K_1$ items. Here $K_1$ is always larger than $K_2$. The RSU aims to expand the set of potential items which might represent the user interest as much as possible. In the next section, we present a novel attention mechanism to model user interest based on the sequence data.

\subsection{Fused Attention Unit\label{sec:fau}}
The FAU decouples the ID field and attribute field and calculate attention scores for adaptive fusion. Although previous decoupling works \cite{liu2021noninvasive,xie2022decoupled} argue that the ID feature which contains important information is more important than the attribute feature, we find that both fields can complement each other when one is of low quality, especially in a long-tail scene.
What's more, we explore the influence of user engagement and try to integrate this information to instruct the model to distinguish the items with deep engagement. 

\subsubsection{Prior Attention Calculation} To calculate the attention score, the original attention score can be formulated as:
\begin{equation}
    att_h = \frac{Q_hK_h^\top}{\sqrt{d}},
\end{equation}
where $Q_h=W_h^QQ$, $K_h=W_h^KK$ and d is the projection matrices dimension to scale the weight. Then h-th head output of the multi-head attention is calculated as:
\begin{equation}
\label{eq:2}
    head_h = \frac{ e^{att_h}}{\sum_{h=1}^H e^{att_h}}V_h,
\end{equation}
where $V_h=W_h^VV$. With integrating multiple types of features, the compound attention score will degrade the performance due to the low rank of attention matrix\cite{xie2022decoupled}. So we separate the features of different domains to calculate the attention score. Then attention score obtained from sequential input can be formulated as:
\begin{equation}
    att_h^{ID} = (R^{(ID)}W_{h(ID)}^Q)(R^{(ID)}W_{h(ID)}^K)^\top / \sqrt{d},
\end{equation}
\begin{equation}
    att_h^{attr} = (R^{(attr)}W_{h(attr)}^Q)(R^{(attr)}W_{h(attr)}^K)^\top / \sqrt{d},
\end{equation}
where ${R^{(ID)}, R^{(attr)}}\in \mathcal{R}^d$ demonstrates the look-up embeddings of ID field and attribute field. The calculated attention score will be fed into the fusion component next.
\subsubsection{Fusion with User Engagement} Given the attention scores, we aim to fuse them according to user engagement. Users always show personalized stickiness on different items. Taking this into consideration, we exploit the engagement gate to strengthen the items which deeply engaged by users for better representation. Then we can obtain the gate formulation as:
\begin{equation}
\label{eq:gate}
    Gate = \sigma(W_2\phi(W_1R^{(ED)})),
\end{equation}
where $R^{(ED)} \in \mathcal{R}^d$, $\sigma$ illustrates the Softmax activation and $\phi$ is the ReLU activation. The gate indicates the importance through historical items on user preference and can be used to adaptively fuse the attention score. Then the fused attention score is formulated as:
\begin{equation}
\label{eq:fusion}
    att_h^{fused} = Gate \odot (\alpha\times att_h^{ID} + (1 -\alpha) \times att_h^{attr}),
\end{equation}
where $\alpha$ denotes the importance of the ID feature in constituting the fused attention score. It is treated as a hyper-parameter to be tuned manually. Noticing that it will stop the gradient of the attribute field with setting $\alpha$ equals 1. $\odot$ is the element-wise product and the deeply engaged items will be selected for user interest extraction. Fusing attention scores with user engagement can not only balance the influence of different field features but also strengthen items with emotional stickiness in historical behaviors.
\subsubsection{Attentive Aggregation} Then we get the alternative formulation of each head in Equation \ref{eq:2}:
\begin{equation}
    head_h = \frac{ e^{att_h^{fused}}}{\sum_{h=1}^H e^{att_h^{fused}}}V_h.
\end{equation}
The outputs of each attention heads of FAU can be concatenated to update the representation of items in behavior sequences. Then we stack a target attention unit following the FAU. In detail, the $att_h$ in target attention is calculated according to the target item and its content feature and the $V_h$ is the feature transformation of the result of FAU. We take this fusion module as the enhancement the representation of target item. Finally, we flatten the sequence embedding and concatenate the user and query embedding for prediction through several fully-connected layers.

\subsection{Model Analysis}
As mentioned in Section 2.2, the bottleneck hinders the extraction of long-term interest for personalized search. So we analyze the key improvements in our model and disclose the detail of how it works. Firstly we discuss the relation with SIM \cite{pi2020search}, a search-based life-long interest model for click-through rate prediction, which shows the underlying equivalence between QIN and SIM. We transfer this recommendation framework and achieve a breakthrough in the modeling of life-long behavior sequences in the search scenario. Then we discuss the connection with the DIF \cite{xie2022decoupled}, which only decouples the side information and ID for sequential recommendation without fusion. It shows that by introducing the user engagement, we can dynamically enhance the attributes' informative influence without the heavy auxiliary attribute predictors.

\subsubsection{Relation with SIM} In \cite{pi2020search}, the authors argue to utilize the complete behaviors for user-friendly service and propose SIM, which searches the relevant subsequences regarding the candidate ad for the lift-long historical behaviors. This work introduces two kinds of relevance score calculation formulation: hard-search and soft-search:
\begin{equation}
    r_i = \left\{
    \begin{aligned}
        &Sign(C_i=C_a) &  & hard-search\\
        &(W_be_i)\odot(W_ae_a)&  &soft-search\\
    \end{aligned}
    \right.
\end{equation}
However, these two units are hard to deploy in search scenarios. The query-driven search engine should not depend as much as possible on the target items, which might lead to embedding drift. What is more serious is that boosted user behaviors without respecting the query can degrade performance and bring a bad experience. In addition, the computational complexity comparison will be elaborated in Section \ref{sec:online} and the effectiveness comparison with the proposed RSU will be shown in Section \ref{sec:rsu}.

Besides, the hard-search method should be utilized sparingly in the content-sharing platform. Unlike e-commerce, the content-sharing platform has a wide variety of items. \emph{Apple} can be categorized as phone or fruit in e-commerce while music, HD pictures, salad even biological science are all likely to be found in content-sharing platforms. That's why we introduce the pre-trained embedding to catch the relevance, and the retrieved subsequences are suitable for modeling user interest in search scene.

\subsubsection{Relation with DIF\label{sec:dif}} In a recent work \cite{xie2022decoupled}, the authors theoretically analyze the expressiveness of attention matrices and move the side information from the input to the attention layer. With multiple predictors, DIF further activates the interaction between side information and item representation. The prediction layer of one kind of attribute fields is formulated as:
\begin{equation}
    y^{attr} = \sigma(W^{attr} s_{attr}^{\top}+b^{attr}),
\end{equation}
where $y^{attr}, W^{attr}$ and $s^{attr}$ denote the probability, learnable parameters and the last element of sequence result of one field separately. The $y^{attr}$ is viewed as the user preference for different attributes activated by $W^{field}$. Recalling Equation (\ref{eq:gate}) (\ref{eq:fusion}), we propose a fusion gate to activate the interaction according to user engagement before the prediction layer. The reasons that we perform engaged activation are two-fold. (1) Comparing with the learnable parameter, user engagement is more explainable about the user's preference for an item.  (2) User engagement encodes high-order feature interactions. The user may be forced to click on a video for many reasons, but after a short thought, called feature interaction in deep learning, he will decide whether to keep watching the video or not. Thus we use it as a gate to guide the model regarding the interest attention and then simplify the prediction layer to avoid overfitting with this powerful feature.

\subsection{Training}

For fair comparison, we closely follow the related work \cite{cheng2022ihgnn} to learn the model parameters. Specifically, given the prediction probability $p$ and ground truth $y$, we optimize the following objective function:
\begin{equation}
    \mathcal{L} = -\sum_{(u,i,q)} y_{uqi} log(p_{uqi})+(1-y_{uqi})log(1-p_{uqi}).
\end{equation}
The $p_{uqi}$ donates the probability of the user $u$ clicking the item $i$ when searching for query $q$. Then we can adopt this probability to train and rank the target items.

\subsection{Online Serving\label{sec:online}}
\begin{figure}[t]
   \centering
  \includegraphics[width=\linewidth]{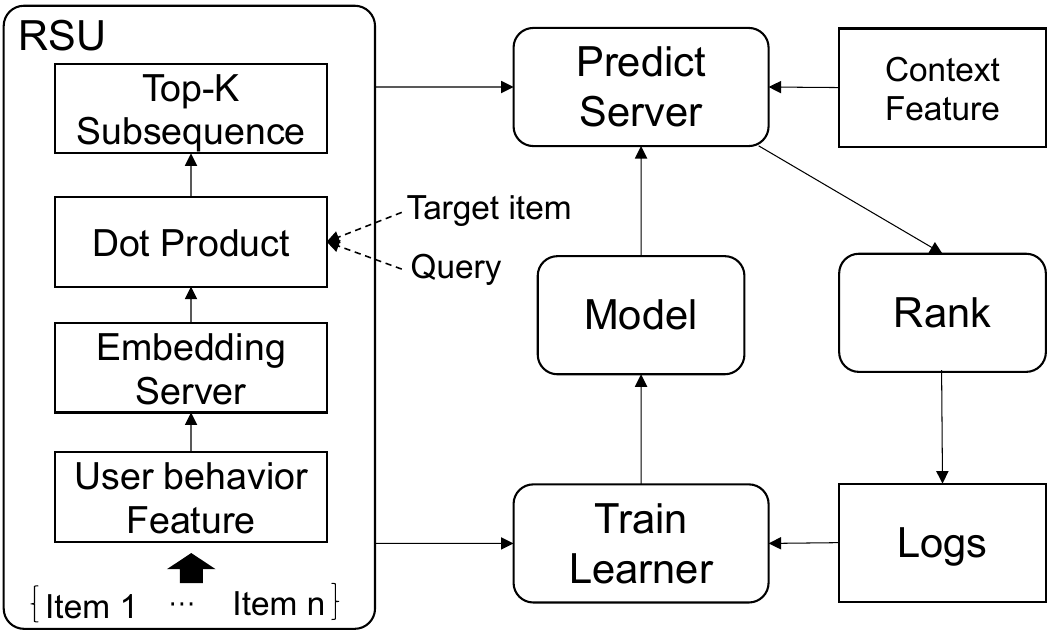}
  \caption{Real-time serving pipelines of training process and inference. Both pipelines join a relevance search unit to seek the effective behaviors with queries and target items from long sequential behavior data.}
  \label{fig:online}
  \Description{}
\end{figure}
In this section, we introduce the pipeline of the implemented real-time video search system. As shown in Figure \ref{fig:online}, without the RSU, the whole pipeline can serve the real-time ctr prediction task in general. The predict server will rank the videos when the user initializes a search request. Then the dumped logs can be used to train the model in some time.

QIN introduces the RSU, to retrieve relevant subsequences with the query based on past behaviors. However, under the same query, completely consistent sequence information cannot distinguish different videos. We further perform two-stage RSU for retrieving the top-K items relevant to the target video. We also tried to experiment directly with a one-stage RSU relevant to target video, but its computational complexity would be a burden for the system. 

Supposed $N, M, D, K_1, K_2$ denote the number of historical behaviors, number of target items, embedding dimension, sequence length of the last unit (if exists) and final sequence length of RSU. The computational complexity of RSU with and without the previous stage can be expressed as $O(ND+MK_1D)+O(Nlog(K_1)+MK_1log(K_2))$ and $O(NMD)+O(NMlog(K_2))$, separately. Obviously, both the second term in the complexity can be ignored because $D$ is far larger than $log(K)$. The simplified formulations can be expressed as $O(ND+MK_1D)$ and $O(NMD)$. For intuitive comparison, we transform $O(NMD)$ equivalently as:
\begin{equation}
    O(NMD) = O(N(1+M-1)D) = O(ND)+O(N(M-1)D).
\end{equation}

With the condition $M\gg1$, the computation complexity is determined by the comparison between $N$ and $K_1$. In fact, the size of target items for Kuaishou can easily reach hundreds, and the system always cache a long period of users' historical behaviors, causing $N\gg K_1$. It is feasible to adopt a two-stage search unit in terms of complexity. 

What's more, the historical behaviors can store up to ten thousand and be duplicated whenever updating. Specially, the deduplication operation will remove about 10\% of redundant items, leaving the one with the most recent server time. These items are usually useless so that we remove them for storing more historical items. The behavioral feature is acquired from the feature center with a Key-Value structure, including the engagement feature, timestamp, author id and so on. The embedding server with the Key-Value data structure, storages the multi-modal embedding \cite{wang2022modality} of all available queries and videos. We have proven that the dot product operation takes the most time, so we speed up with the parallel paradigm.

\section{EXPERIMENTS}
In this section, we conduct experiments to validate the effectiveness of the proposed method QIN. Our experiments are intended to answer the following research questions:
\begin{itemize}
    \item \textbf{RQ1}: Does QIN outperform state-of-the-art PSR methods?
    \item \textbf{RQ2}: How do different components (i.e., RSU and FAU) affect QIN performance?
    \item \textbf{RQ3}: How does $\alpha$ balance the influence of both fields?
    \item \textbf{RQ4}: Does QIN beat the highly-optimized production baseline in real world system? 
\end{itemize}
\subsection{Experimental Settings}
\subsubsection{Dataset} Our experiments are conducted on Amazon\footnote{http://jmcauley.ucsd.edu/data/amazon/} dataset\cite{mcauley2015image}. We select four subsets of Amazon dataset, which are \emph{Amazon Beauty, CDs \& Vinyl, Electronics} and \emph{Kindle Store}. Van Gysel et al. \cite{van2016learning} introduced a method to extract queries from the categories. Many recent pieces of approaches \cite{ai2019zero, ai2017learning, bi2020transformer, cheng2022ihgnn} follow this synthetic process and view it as a common benchmark dataset. To reduce the experiment workload and make the comparison fair, we generate several semi-synthetic datasets in this way. 5-core filtering has been conducted and the sequence length is set to 10. User engagement is extracted from the review information such as the overall rate and helpful rate. Besides, the overall behaviors are all the interactions of users and the search behaviors are the interactions of user-query groups. 

\begin{table}[t]
\caption{Dataset Statistics}
    \centering
    \begin{tabular}{ccccc}
    \toprule
         &users&queries&items&interactions\\
         \midrule
         Beauty&22,363&251&12,101&198,502\\
         \midrule
         CDs \& Vinyl&75,258&22,204&64,443&1,097,592\\
         \midrule
         Electronics&192,403&1,386&63,001&1,689,188\\
        \midrule
         Kindle Store&68,223&25,416&61,934&982,619\\
    \bottomrule
    \end{tabular}
    \label{tab:data}
\end{table}
\begin{table*}[t]
  \caption{Performance Comparison with the state-of-the-art methods.}
  \label{tab:rq1}
  \begin{tabular}{c|ccc|ccc|ccc|ccc}
        \toprule
        Dataset & \multicolumn{3}{c}{Beauty} & \multicolumn{3}{c}{CDs \& Vinyl}& \multicolumn{3}{c}{Electronics}& \multicolumn{3}{c}{Kindle Store}\\
        \midrule 
        Metric@4 & NDCG & MRR & HR & NDCG & MRR & HR& NDCG & MRR & HR& NDCG & MRR & HR\\
        \midrule
        HEM         &0.0655 &0.0540	&0.0999 &0.0619 &0.0506 &0.0956 &0.0655	&0.0537 &0.1009 &0.0405	&0.0328	&0.0638\\
        ZAM         &0.1269	&0.1082	&0.1826 &0.1453	&0.1242	&0.2083 &0.1861	&0.1577	&0.2710 &0.1244	&0.106	&0.1794\\
        TEM         &0.1370	&0.1175	&0.1953 &0.1800	&0.1584	&0.2445 &0.1967	&0.1710	&0.2735 &0.1723	&0.1481	&0.2443\\
        GraphSRRL   &0.0826 &0.0688	&0.1238 &0.1286	&0.1069	&0.1933 &0.2335	&0.2034	&0.3233&0.1024	&0.08707&0.1482\\
        IHGNN       &0.1038	&0.0871 &0.1539	&0.2066	&0.1819	&0.2804 &0.2408	&0.214	&0.3205&0.1435	&0.1246	&0.1999\\
        MultiResAttn&0.1806 &0.1547 &0.2579 &0.2981 &0.2646 &0.3979 &0.3251 &0.2919 &0.4239&0.2225	&0.1914	&0.3154\\
        \midrule
        QIN        &\textbf{0.2305}  &\textbf{0.2002} &\textbf{0.3206} &\textbf{0.3693} &\textbf{0.3301} &\textbf{0.4854} &\textbf{0.3935} &\textbf{0.3573} &\textbf{0.5011}& \textbf{0.2538}	&\textbf{0.2218}	&\textbf{0.349}\\
        \bottomrule
    \end{tabular}
\end{table*}
\subsubsection{Compared Methods} The following methods are tested in our experiments:
\begin{itemize}
    \item \textbf{HEM} \cite{ai2017learning} jointly learns the representations of users, products and queries in one latent vector space and this is one of the initialized research in personalized product search.
    \item \textbf{ZAM} \cite{ai2019zero} extends HEM to fuse the user representation and purchased product sequence with attention structure to make a personalized prediction.
    \item \textbf{TEM} \cite{bi2020transformer} encodes the query and purchased products sequence with a transformer-based architecture \cite{vaswani2017attention}. 
    \item \textbf{GraphSRRL} \cite{liu2020structural} propose three conjunctive graph patterns in users-queries-products interactive graph to learn the the structural relationship.
    \item \textbf{IHGNN} \cite{cheng2022ihgnn} models the collaborative signal among users, queries and products based on the constructed interaction hypergraph. It makes a huge gain on GraphSRRL so we compare IHGNN as the SOTA method for search ranking.
    \item \textbf{MultiResAttn} \cite{kocayusufoglu2022multi} exploits higher-order temporal dependencies between users’ search queries and item history and proposes a novel attention module to capture users’ interests within designated temporal resolutions. Although it does not compare the above baselines, we reproduce this method as the competitive method for search ranking.
\end{itemize}

Besides, for the in-depth ablation studies, we further choose some competitive approaches for sequential recommendation to validate the effectiveness of the proposed FAU stated in Section \ref{sec:fau}. 
\begin{itemize}
    \item \textbf{DIN} \cite{zhou2018deep} designs a local activation unit to focus on the related user behaviors. We reproduce the unit as the method of sequence pooling to replace the FAU for the ablation study.
    \item \textbf{SASRec} \cite{kang2018self} flexibly tends to consider dependencies on datasets of different densities. We extract the self-attentive block for validation.
    \item \textbf{DIF} \cite{xie2022decoupled} decouples the side information and item ID for the higher rank attention matrices and adaptive gradients. We have stated the relationship between DIF and our attention module in Section \ref{sec:dif}. Experimental results show the superiority of our proposed FAU.
\end{itemize}

\subsubsection{Evaluation} In our experiments, for all Amazon datasets, we use the \emph{leave-one-out} strategy for evaluation. Specifically, for each user-item interaction, the last two items are reserved as validation and testing data, respectively, and the rest are utilized for training models. The best validation checkpoint will be used to show the test performance. The performance of models is evaluated by top-N Normalized Discounted Cumulative Gain (NDCG@N) and top-N Hit ratio (HR@N), top-N Mean reciprocal rank (MRR@N) with N chosen from \{4,8,20\}. We use the comparison results of the \textbf{top 4} metrics as the core reference.

\subsubsection{Implementation Details} We implement our QIN with TensorFlow. The embedding size is set to 8. We randomly sample 100 negative items for each user-query-item interaction. We optimize QIN by the Adam method with a default learning rate of 0.001 and a default batch size of 512 for all the models unless the authors specially state it. The embeddings are initialized by using the default Xavier uniform method. Models are trained on multiple 78-core CPUs and we speed up with Horovod\footnote{https://github.com/horovod/horovod}.

\begin{table}[t]
  \caption{The performance lift of QIN w.r.t. different N on four datasets.}
  \label{tab:overall}
   \resizebox{\columnwidth}{!}{
      \begin{tabular}{c|cc|cc|cc}
        \toprule
        \multirow{2}{*}{Method}&\multicolumn{2}{c}{Metric@4}&\multicolumn{2}{c}{Metric@8}&\multicolumn{2}{c}{Metric@20}\\
        ~&NDCG&MRR& NDCG&MRR& NDCG&MRR\\
        \bottomrule
        \multicolumn{5}{l}{Beauty}\\
        \hline
        MultiResAttn  &0.1806 & 0.1547 & 0.223 & 0.1744&0.2716&0.1887 \\
        QIN    &\textbf{0.2305 }&\textbf{0.2002 }&\textbf{0.2765 }&\textbf{0.2216 }&\textbf{0.3285}&\textbf{0.237}\\
        $\Delta$ \%    &27.63&29.41&23.99&27.06&20.95&25.6\\
        \bottomrule
        \multicolumn{5}{l}{CDs \& Vinyl}\\
        \hline
        MultiResAttn  &0.2981 & 0.2646 & 0.3499 & 0.2886 &0.4026&0.3043\\
        QIN    &\textbf{0.3693 }&\textbf{ 0.3301 }&\textbf{ 0.4234 }&\textbf{ 0.3553}&\textbf{0.4775}&\textbf{0.3715}\\
        $\Delta$ \%    &23.88&24.75&21.0&23.11&18.6&22.08 \\
        \bottomrule
        \multicolumn{5}{l}{Electronics}\\
        \hline
        MultiResAttn  &0.3251 &0.2919 &0.3714 &0.3133 &0.4181&0.3272\\
        QIN    &\textbf{0.3935 }&\textbf{ 0.3573 }&\textbf{ 0.4419 }&\textbf{ 0.3797}&\textbf{0.4908}&\textbf{0.3944}\\
        $\Delta$ \%    &21.04&22.4&18.98&21.19&17.39&20.54\\
        \bottomrule
        \multicolumn{5}{l}{Kindle Store}\\
        \hline
        MultiResAttn  &0.2225 & 0.1914 & 0.2774 & 0.2161 &0.3331&0.2333\\
        QIN    &\textbf{0.2538 }&\textbf{ 0.2218 }&\textbf{ 0.306 }&\textbf{ 0.246}&\textbf{0.3674}&\textbf{0.2642}\\
        $\Delta$ \%    &14.07&15.88&11.51&13.82&10.31&13.24\\
        
        \bottomrule
      \end{tabular}
    }
\end{table}

\subsection{Performance Comparison (RQ1)}
Table \ref{tab:rq1} shows the performance comparison with advanced methods. Based on these results, we can see that: 
\begin{itemize}
    \item For three basic PSR baselines, QIN outperforms other methods by a large margin, while TEM and ZAM are better than HEM, demonstrating the superiority of attention-based methods on sequence data for search ranking.
    \item By comparing the graph-based methods with other methods, we find that the graph-based methods are not so good. The reason is likely to be that we follow \cite{van2016learning} to make a random mask for the training query, and the graph-based methods degrade a lot when there are many new nodes in the graph.
    \item Among the baselines, MultiResAttn exhibits the strongest performance, which is better than IHGNN and TEM. The improvement on baselines proves the superiority of QIN and QIN performs an average of \textbf{21.19\%} better in terms of all metrics on the four datasets. 
\end{itemize}

We also perform the relative gain of Metric@8, Metric@20 between QIN and MultiResAttn for all the datasets in Table \ref{tab:overall}. In conclusion, QIN performs the consitent improment and does not degrade much with a various N.

\subsection{Ablation Studies (RQ2)}
\begin{table}[t]
  \caption{Ablation study of RSU method in our model.}
  \label{tab:rsu}
  \resizebox{\columnwidth}{!}{
  \begin{tabular}{c|ccc|ccc}
        \toprule
        Dataset  & \multicolumn{3}{c}{Beauty}& \multicolumn{3}{c}{CDs \& Vinyl}\\
        \cmidrule{1-7}
        Method & NDCG & MRR & HR & NDCG & MRR & HR\\
        \midrule
        $RSU_{one}$         &0.1683	&0.1446	&0.2388 &0.2551	&0.2248	&0.3453 \\
        $RSU_{SIM}$     &0.206	&0.1782	&0.289 &0.3098	&0.2741	&0.4161 \\
        $RSU_{two}$   &\textbf{0.259}	&\textbf{0.2274}&\textbf{0.3531} &0.3054	&0.2699	&0.4109 \\
        QIN        &0.2305	&0.2002	&0.3206&\textbf{0.3693} &\textbf{0.3301} &\textbf{0.4854} \\
        \bottomrule
    \end{tabular}
    }
\end{table}
In order to figure out the contribution of different components in our proposed QIN, we conduct an ablation study for each proposed component on Beauty and CDs \& Vinyl datasets. The ablation study can prove the sig. The results on Electronics and Kindle are omitted due to the consistency of conclusions.
\subsubsection{Impact of RSU \label{sec:rsu}} Table \ref{tab:rsu} shows the results of QIN and its variants. We perform the proposed QIN with one-stage RSU relevant to the query, QIN with one-stage RSU relevant to the target item and QIN with two-stage RSU, called $RSU_{one}$, $RSU_{SIM}$, $RSU_{two}$. QIN means that we keep the results of $RSU_{one}$ and $RSU_{two}$ to learn the user preference and $RSU_{SIM}$ is reproduced according to the search unit in \cite{pi2020search}. We have three main observations:
\begin{itemize}
    \item Focusing on $RSU_{one}$, the performance degrades than QIN as expected. Because it only uses the query to search the user's behavior history, the subsequence from the RSU will cover a relatively large range of items, which might be irrelevant to the target item. However, the $RSU_{one}$ still outperforms most baselines except MultiResAttn in same experimental setting. This phenomenon indicates the effectiveness of the insight for relevance search.
    \item Comparing with $RSU_{SIM}$, which we discuss their relation in Section \ref{sec:dif}, QIN achieves an improvement by prepending a search unit with queries before retrieving user behavior with the target item, which is coincident with our intuition. Although the subsequence only retrieved by the target item provide useful personalized preferences, the burden for online serving make this method undeployable.
    \item Focusing on the inconsistency of performance lift of QIN on different datasets. Comparing with $RSU_{two}$, QIN consistently outperforms $RSU_{two}$ on CDs \& Vinyl dataste, but not on Beauty dataset. Regarding this phenomenon, it should be pointed out: As shown in Table \ref{tab:data}, the CDs \& Vinyl dataset has a far larger number of queries than the Beauty dataset, leading to a very large distribution gap between the two datasets. Actually, queries comes to a long-tail distribution in the real world and the average interactions per query are more likely to the CDs \& Vinyl dataset. As such, we eventually retain the first stage for QIN.
\end{itemize}
\begin{table}[t]
  \caption{Ablation study of FAU method in our model.}
  \label{tab:fau}
  \resizebox{\columnwidth}{!}{
  \begin{tabular}{c|ccc|ccc}
        \toprule
        Dataset  & \multicolumn{3}{c}{Beauty}& \multicolumn{3}{c}{CDs \& Vinyl}\\
        \cmidrule{1-7}
        Method & NDCG & MRR & HR & NDCG & MRR & HR\\
        \midrule
        $MEAN$  &0.1208	&0.1009	&0.183&0.161 &0.1398 &0.2242 \\
        $DIN$   &0.1456&0.1248&0.2078&0.2018&0.1761	&0.2782\\
        $SASRec$&0.1759	&0.1507	&0.2512	&0.2786	&0.2434	&0.3832 \\
        $DIF$   &0.2147	&0.1892	&0.2906&0.3079	&0.2731	&0.4115 \\
        $QIN_{s}$&\textbf{0.2478}&\textbf{0.2176}&\textbf{0.3379}&0.3225&0.2865&0.4298 \\
        $QIN_{id}$&0.1498&0.1286&0.2133&0.2741&0.2416&0.371 \\
        $QIN$   &0.2305	&0.2002&0.3206&\textbf{0.3693} &\textbf{0.3301} &\textbf{0.4854} \\
        \bottomrule
    \end{tabular}
    }
\end{table}
\subsubsection{Impact of FAU} In QIN, we employ a novel attention module on the behavior sequence. For a fair comparison, we reproduce some competing pooling methods combined with the result of RSU and explore serval different choices on the engagement fusion. As shown in Table \ref{tab:fau}, $QIN_{s}$ means that the engagement gate is a scalar and $\sigma$ in Equation \ref{eq:gate} illustrates Sigmoid activation. We also follow \cite{liu2021noninvasive} to set the $V_h$ in Equation \ref{eq:2} as the feature transformation of ID embedding, called $QIN_{id}$. Method \emph{MEAN} denotes the mean pooling of sequential embedding. We have the following observations:
\begin{itemize}
    \item The best setting in general is using user engagement as a gate to activate historical behaviors. The performance of $QIN_{s}$ is closed to $QIN$. Both $QIN$ and $QIN_{s}$ outperform other methods, which shows that the fused attention unit is effective. We choose the form of vector as the default engagement gate because the scalar might miss information after the Sigmoid activation.
    \item DIF shows the second-best performance among different attention modules. The main difference between DIF and FAU is whether to utilize user engagement. This phenomenon proves the effectiveness of the insight to fuse historical behaviors with user engagement.
    \item $QIN_{id}$ degrades much since it ignores the rich content feature. It demonstrates that the attribute feature and ID feature can complement each other in the long-tail scene, and we should combine them for a better item representation.
\end{itemize}

\subsubsection{Sensitive of embedding size and hidden units}
Considering the embedding size might affect the overall performance of the proposed model. We conduct several couples of experiments with the embedding size chosen from \{8,16,32\} and the last few FC layers setting in \{[16,8,1], [128,64,1]\}, which named $FC_1$, $FC_2$ for short. We show the experimental result in Table \ref{tab:sensitive}. 
 With the larger embedding size, there is not always an improvement for QIN. This paper sets the embedding size to 8 for the reason to reduce the workload and it is shown enough to prove the effectiveness of the experiment. Tuning the hidden units of FC layers will work if the embedding size is 16 or 32. Reviewing the performance, it may be due to it's just tricky for performance lift. Thus we set the FC layers to [16,8,1] by default and try the best performance no longer.
\subsubsection{Sensitive of sequence length}
\begin{figure}[t]
   \centering
  \includegraphics[width=\linewidth]{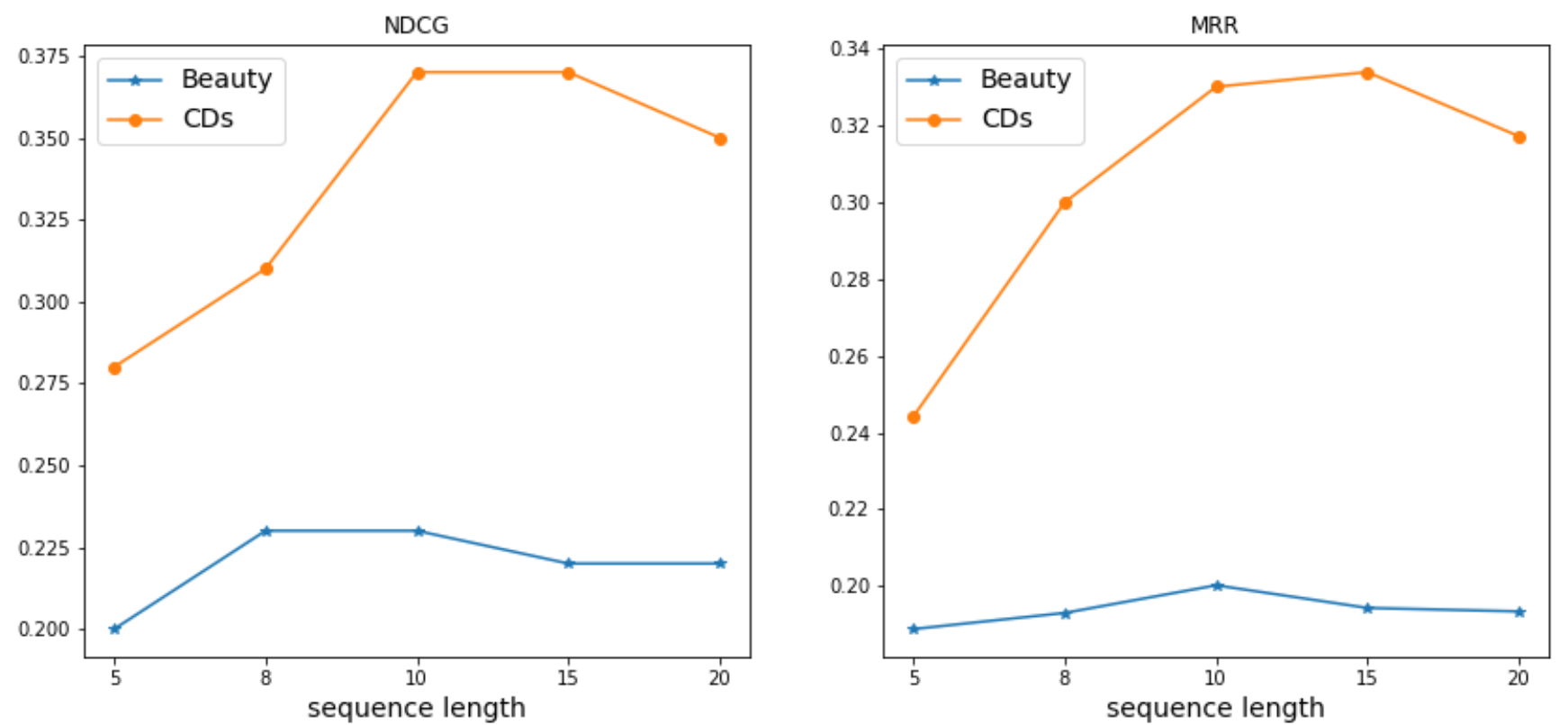}
  \caption{Performance of QIN w.r.t. different sequence length on Beauty and CDs \& Vinyl datasets.}
  \label{fig:len}
  \Description{}
\end{figure}

We further try different setting for sequence length. As shown in Figure \ref{fig:len}, QIN degrades with large sequence length. This is because the soft search process in RSU will introduces noise when the sequence length is larger than the number of user historical behaviors.
\balance
\begin{table}[t]
  \caption{The impact of embedding size and hidden units. The Best results are marked bold and the second best are underlined. }
  \label{tab:sensitive}
  \resizebox{\columnwidth}{!}{
  \begin{tabular}{c|ccc|ccc}
        \toprule
        Dataset  & \multicolumn{3}{c}{Beauty}& \multicolumn{3}{c}{CDs \& Vinyl}\\
        \cmidrule{1-7}
        Emb/FC & NDCG & MRR & HR & NDCG & MRR & HR\\
        \midrule
        $8/FC_1$  &0.1282	&0.1099	&0.183&\underline{0.3693}&\underline{0.3301}	&\underline{0.4854} \\
        $16/FC_1$   &\underline{0.2524}&\underline{0.2221}&\underline{0.3429}&0.3604&0.323	&0.4715\\
        $32/FC_1$       &0.2032&0.175&0.2873&0.367&0.3285	&0.4814 \\
        \midrule
        $8/FC_2$   &0.2462&0.2148&0.3396&0.3461&0.308&0.4598 \\
        $16/FC_2$        &0.241&\textbf{0.274}&\textbf{0.3722}&\textbf{0.3784}&\textbf{0.3394}&\textbf{0.4941} \\
        $32/FC_2$       &\textbf{0.2677}&0.2347&0.3657&0.3736&0.3343&0.4904 \\
        \bottomrule
    \end{tabular}
    }
\end{table}
\begin{figure}[t]
   \centering
  \includegraphics[width=\linewidth]{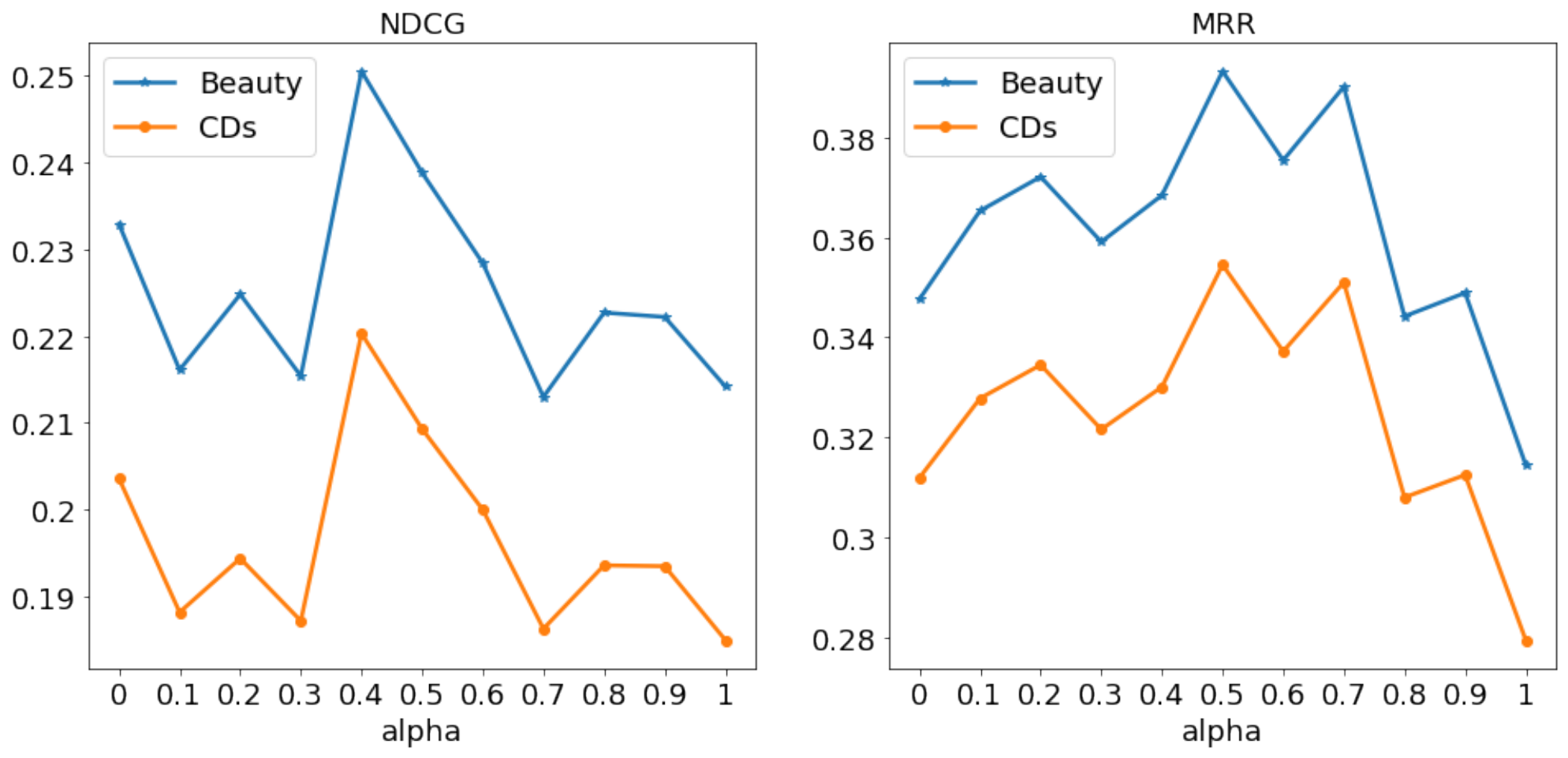}
  \caption{Performance of QIN w.r.t. different $\alpha$ on Beauty and CDs \& Vinyl datasets.}
  \label{fig:hyper}
  \Description{}
\end{figure}

\subsection{Hyper-parameter Studies (RQ3)}

The most important hyper-parameter to tune in QIN is $\alpha$. As stated in Equation \ref{eq:2}, $\alpha$ balance the influence of the ID feature and attribute feature. As shown in Figure \ref{fig:hyper}, QIN reaches the best performance at around 0.5, which admits the same influence of both fields. Noting that even in extreme cases (e.g. setting $\alpha$ 1 or 0), QIN can still maintain a good effect. This proves that the proposed model will not rely too much on ID features or attribute features, making it easier to deal with the long-tail situations. Especially when the user's behavior is relatively sparse, our model will not degrade too much.

\subsection{Online Evaluation (RQ4)}

\textbf{Online A/B Testing}. 
We also report on the online A/B testing, which compared the performance of QIN to other deep learning models on the task of ctr prediction for one month. As shown in Table \ref{tab:ab}, we evaluate the performance by observing the lift in three metrics for one month, involving more than 100M queries. The metric of user engagement is efficient viewing and query change rate. Efficient viewing measures the video that has been played over 4 seconds, and the query change rate indicates whether a user will initiate the search request with another query, the smaller the better. The performance lift of ctr has reached \textbf{7.6\%}, which represents a significant improvement for online experiments. Now QIN has served the main traffic every day. 
\begin{table}
  \caption{QIN's improvement for online A/B test. Query change rate is desired to be small while ctr and efficient view is large.}
  \label{tab:ab}
  \begin{tabular}{cccc}
    \toprule
    Metric&ctr &efficient view& query change rate\\
    \midrule
    Lfit rate&+7.6\%&+24.1\%&-7.2\%\\
  \bottomrule
\end{tabular}
\end{table}

\section{CONCLUSION AND FUTURE WORK}
In this work, we argued the user interest model based on search behaviors learn the representation with bottleneck and analyzed the limitation with real-world data. To address the problem, we proposed QIN which consists of two units --- relevance search unit and fused attention unit. In the RSU, we seek the subsequences from the raw complete user behaviors relevant to the query. Then the subsequences are fed into another search unit to retrieve the top-K sub-subsequences relevant to the target item. In the FAU, we decoupled the ID field and attribute field to calculate the attention scores separately, and adopted user engagement to guide the attention scores' fusion of both fields. We also conduct experiments and A/B testing to demonstrate the strengths of QIN. 

We believe the insights of QIN are inspirational to future developments of the user interest model for personalized search ranking. In future work, we plan to explore instant interest for the search, which is benefit for the real time prediction. In this way, the negative feedback will be introduced for better user engagement.

%



\bibliography{cite}
\end{document}